%% file: Master.tex
\documentclass[10pt,twoside,twocolumn]{book}
\usepackage{jrjc}
\usepackage{soul}	% Pour barrer du texte

\begin{document}

%%%%%%%%%%%%%%%%%%%%%%%%%%%%%%%%%%%%%%%%%%%%%%%%%%%%%%%%%%%%

%\rjcinclude{session}{Laviron}

%
  {%                            in local group to allow private
   %                            redefinitions
    \cleardoublepage%           make sure we start on the right...
    \subimport{session/Laviron/}{Laviron}
  }
  % if you do not want to include a photo of you

%%%%%%%%%%%%%%%%%%%%%%%%%%%%%%%%%%%%%%%%%%%%%%%%%%%%%%%%%%%%

\end{document}

%% file: Laviron.tex
%%%%%%%%%%%%%%%%%%%%%%%%%%%%%%%%%%%%%%%%%%%%%%%%%%%%%%%%%%%%%%%%%%%%%%%%%%%%%%%%%%%%%%%%%%%%%%%%%%%%%%%%%%%%%%%%%%%%%%%%%%%%%%%%%
%                                                                                                                               %
%                                  Proceedings of Journée de Rencontre des Jeunes Chercheurs 2019                               %
%                                   Adrien LAVIRON - Université Paris-Sud/CSNSM CNRS IN2P3                                      %
%                                                                                                                               %
%%%%%%%%%%%%%%%%%%%%%%%%%%%%%%%%%%%%%%%%%%%%%%%%%%%%%%%%%%%%%%%%%%%%%%%%%%%%%%%%%%%%%%%%%%%%%%%%%%%%%%%%%%%%%%%%%%%%%%%%%%%%%%%%%

%\def \vt{\color{magenta}}
\def \vt{}

\title{Development of an advanced Compton telescope for MeV--range gamma--ray astronomy}
\author{\large{Adrien LAVIRON}, \normalsize{V. Gourlaouen, C. Hamadache, C. Hiver, J. Kiener, J. Peyre, V. Tatischeff}}
\institute{Universit\'e Paris-Sud, CNRS/IN2P3, CSNSM, 91405 Orsay, France}

\abstract{
%\lipsum[1]
An advanced Compton telescope appears to be the best instrument concept for the next generation gamma--ray space observatory in the MeV range. A first prototype of advanced Compton telescope is being developed to match the constraints of a nano satellite mission, with the scientific objective of measuring gamma--ray burst prompt emission polarization. Instrumental developments at CSNSM for this project are focusing on the position--sensitive calorimeter module, made of a monolithic inorganic CeBr$_3$ scintillator read by a pixelated photodetector. %{\vt \st{Spectral resolution down to $5.1\%$ at $662\,\mathrm{keV}$ have been attained}}.
3D position reconstruction is obtained by deep--learning algorithms that have been optimized down to an uncertainty of $2\,\mathrm{mm}$ for each spatial direction.
}

\maketitle

%\footnotetext{Formerly University Paris-Sud, CNRS/IN2P3, CSNSM, 91405 Orsay, France}
\section*{Introduction}

MeV--range gamma--ray astronomy contains a large number of science themes such as the nucleosynthesis and chemical
evolution of the universe, cosmic rays physics, or %astrophysical jets and the 
multi--messenger astronomy. The best concept for next generation instruments is thought to be a space--borne advanced Compton imager.

Compton imaging relies on the dominant process of interaction of gamma rays with matter in this energy range, 
Compton scattering. In a next generation MeV--range gamma--ray space telescope as proposed in~\cite{eastrogam}, 
the incoming photon undergoes an inelastic scattering in one or several layers of position--sensitive silicon strip detector 
before being absorbed in a position--sensitive calorimeter based on inorganic scintillators. 
The measurement of both positions and energy deposits enables the determination of the photon's source direction~\cite{zoglauer}.
It also enables measurement of the linear polarization of the incident gamma rays, which can provide a powerful diagnostic of the emission processes.

These measurements may be performed either in a large space observatory with a modular design or with a fleet of nano--satellites that enables all--time full--sky coverage, preferable for studying transient phenomena such as gamma--ray bursts.

\section*{Current instrumental developments}

The main task of the CSNSM research team on this project is to develop a calorimeter module, that could be used either as a calorimeter in a nano--satellite or as a part of a large observatory's calorimeter. This study follows the work of~\cite{2015NIMPA.787..140G} and~\cite{2016NIMPA.832...24G}.

\begin{figure}[htbp]
\begin{center}
\includegraphics[width=.9\linewidth,keepaspectratio]{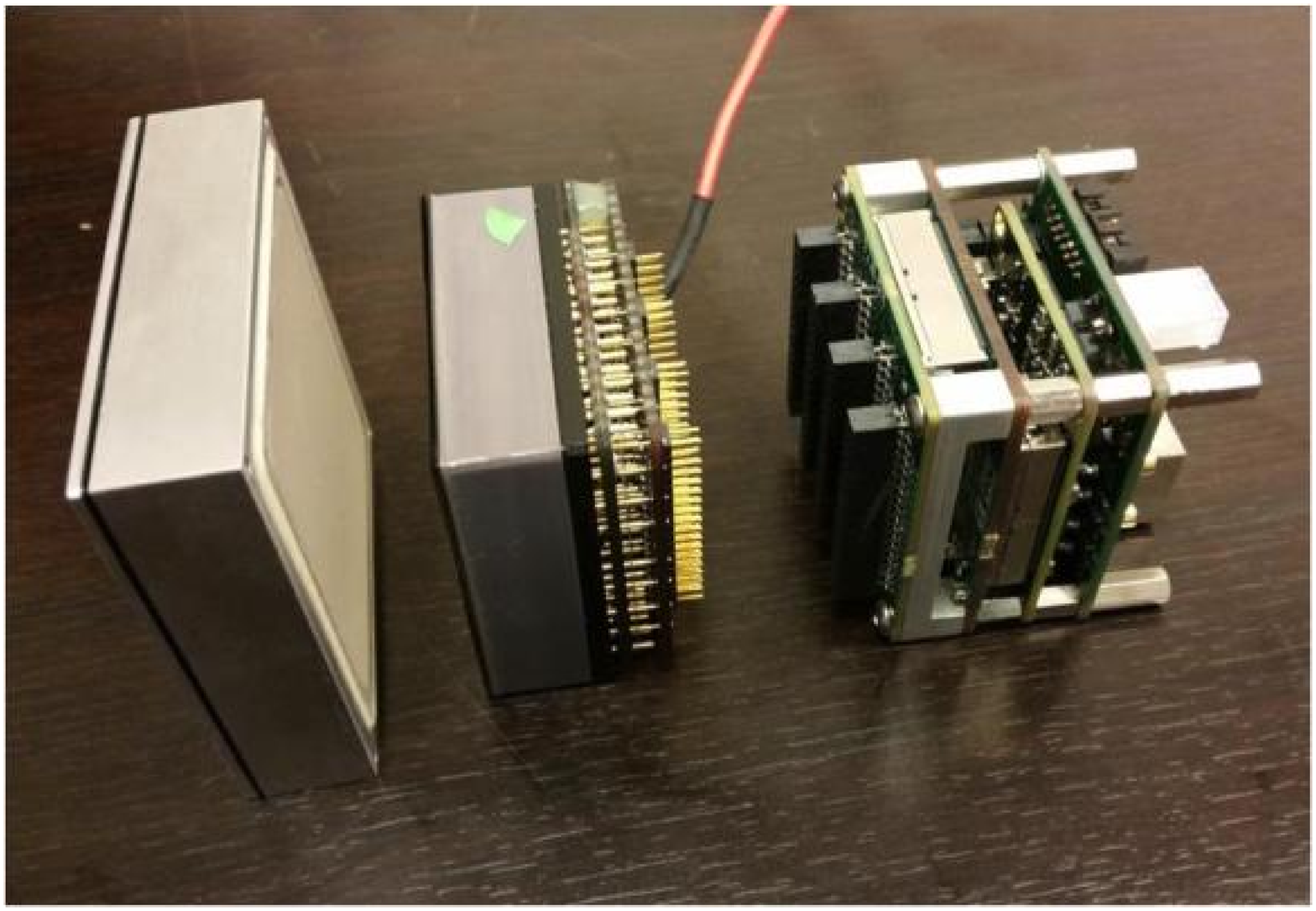}
\end{center}
\caption{Photograph of the three parts of a module. From left to right : Hygroscopic CeBr$_3$ scintillator in its aluminium casing, Hamamatsu MAPMT, and ROSMAP front--end electronics.}
\label{fig:module}
\end{figure}

The selected design for this calorimeter module is a monolithic square $51 \times 51 \times 10$ mm Cerium Bromide (CeBr$_3$) 
inorganic hygroscopic crystal (manufactured by {\vt SCIONIX} company) read by a 64 channels multi--anode photomultiplier tube (MAPMT) or by an array of $8\times8$ silicon photomultipliers (SiPM). Those pixelated photodetectors are read by a commercial, self--triggered electronic system ROSMAP (designed by {\vt IDEAS} company). Such a module can be seen on Figure~\ref{fig:module}. 

\begin{figure}[t!]
\begin{center}
\includegraphics[width=.9\linewidth,keepaspectratio]{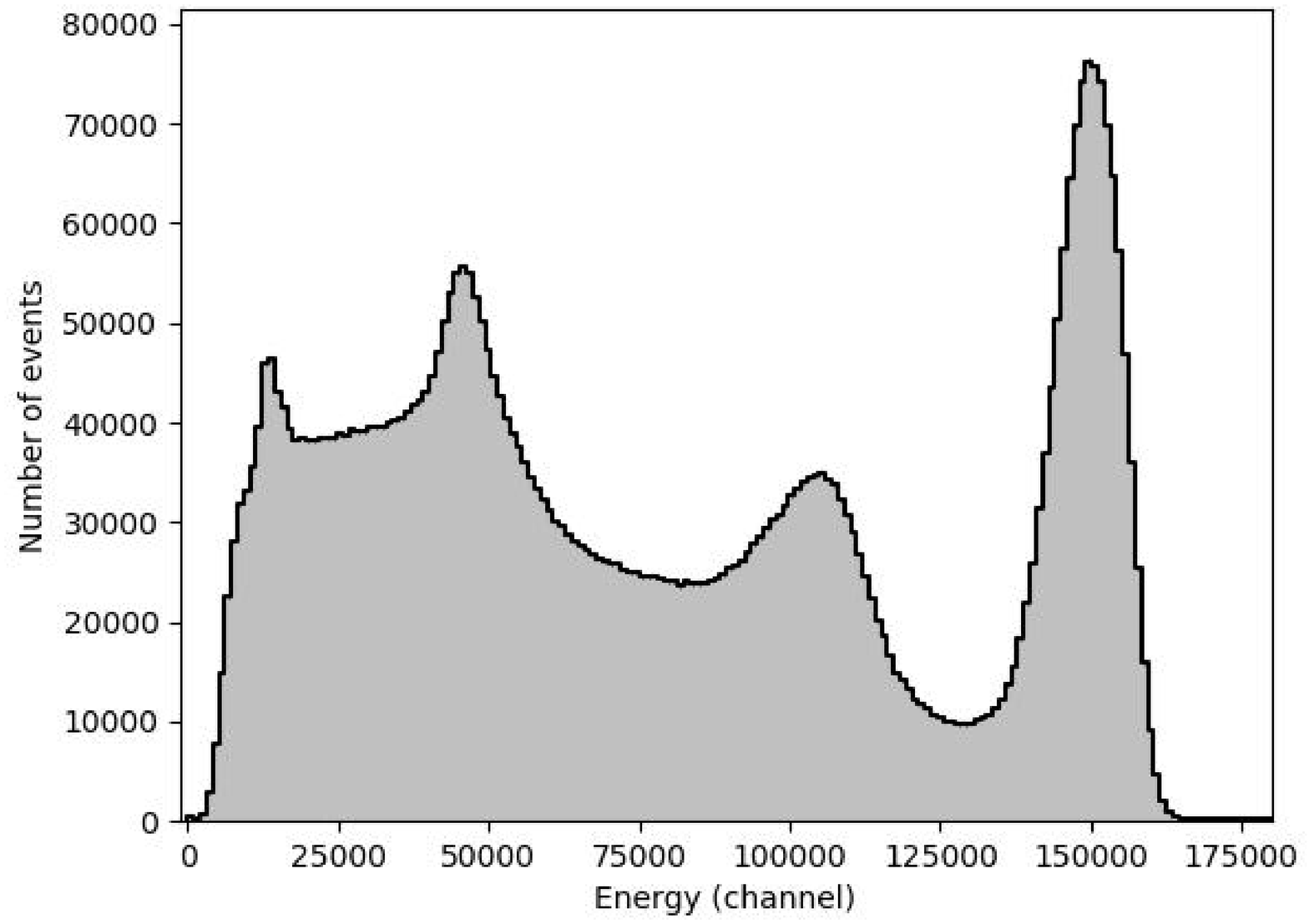}
\end{center}
\caption{Spectrum of a ${}^{137}$Cs radioactive source ($662\,\mathrm{keV}$ line centered on channel $\approx 150000$)}
\label{fig:spectrum}
\end{figure}

In case of interaction of a gamma ray in the crystal, this scintillator produces an amount of light proportional to the energy deposit. This light is collected by the pixelated photodetector, and the sum of the signals read for each pixel enables energy measurements. Figure~\ref{fig:spectrum} shows an energy spectrum of a ${}^{137}$Cs radioactive source acquired with this module. 

\begin{figure}[htbp]
\begin{center}
\includegraphics[width=.9\linewidth,keepaspectratio]{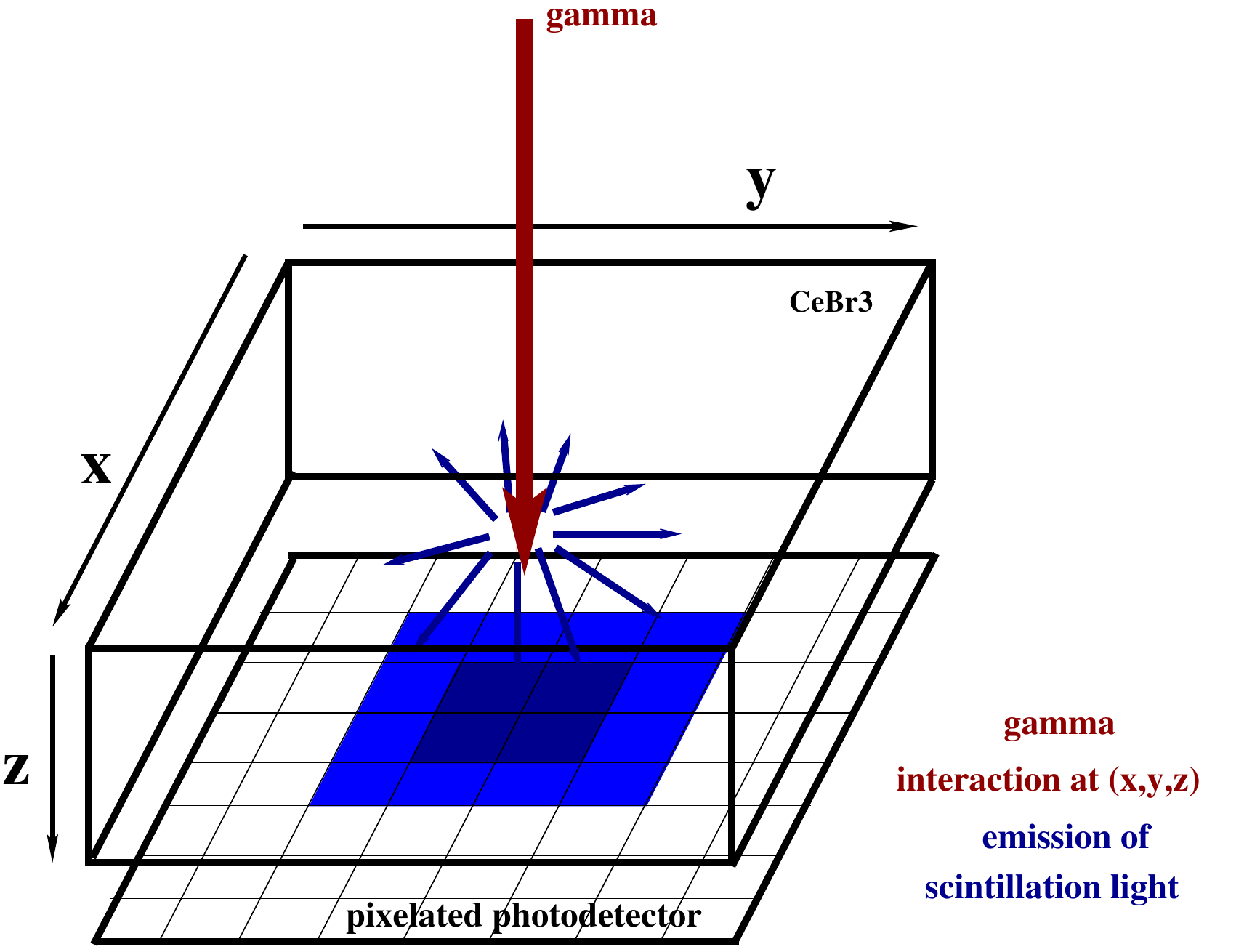}
\end{center}
\caption{Schema of the calorimeter module. The gamma ray interacts in the CeBr$_3$ that emits scintillation light. This light is detected by a pixelated photodetector optically coupled to the scintillating crystal to analyze the shape of the scintillation light distribution.}
\label{fig:principle}
\end{figure}

The use of a pixelated photodetector also enables the analysis of the shape of the scintillation light distribution, in order to reconstruct the 3D position of first interaction of the gamma ray in the calorimeter. This principle is illustrated on Figure~\ref{fig:principle}. 

\section*{Data analysis}

To assess the position reconstruction ability of the module, we have {\vt developed} a test bench {\vt consisting} of a collimated radioactive source generating a gamma--ray beam of width $2\,\mathrm{mm}$, and a module that can be automatically moved along two axes perpendicular to the direction of the gamma--ray beam. The source can irradiate either the $xy$ or the $yz$ plane of the scintillator {\vt (see Fig.~\ref{fig:principle})}. That way, either the $(x,y)$ coordinates or the $(y,z)$ coordinates of the gamma--ray first interaction are known.

In this part, we will briefly discuss the morphology of the events, before presenting the deep--learning algorithms that we use to reconstruct the position first in the $xy$ plane and then in 3D.

\subsection*{Events morphologies}

\begin{figure}[htbp]
\begin{center}
\includegraphics[width=.9\linewidth,keepaspectratio]{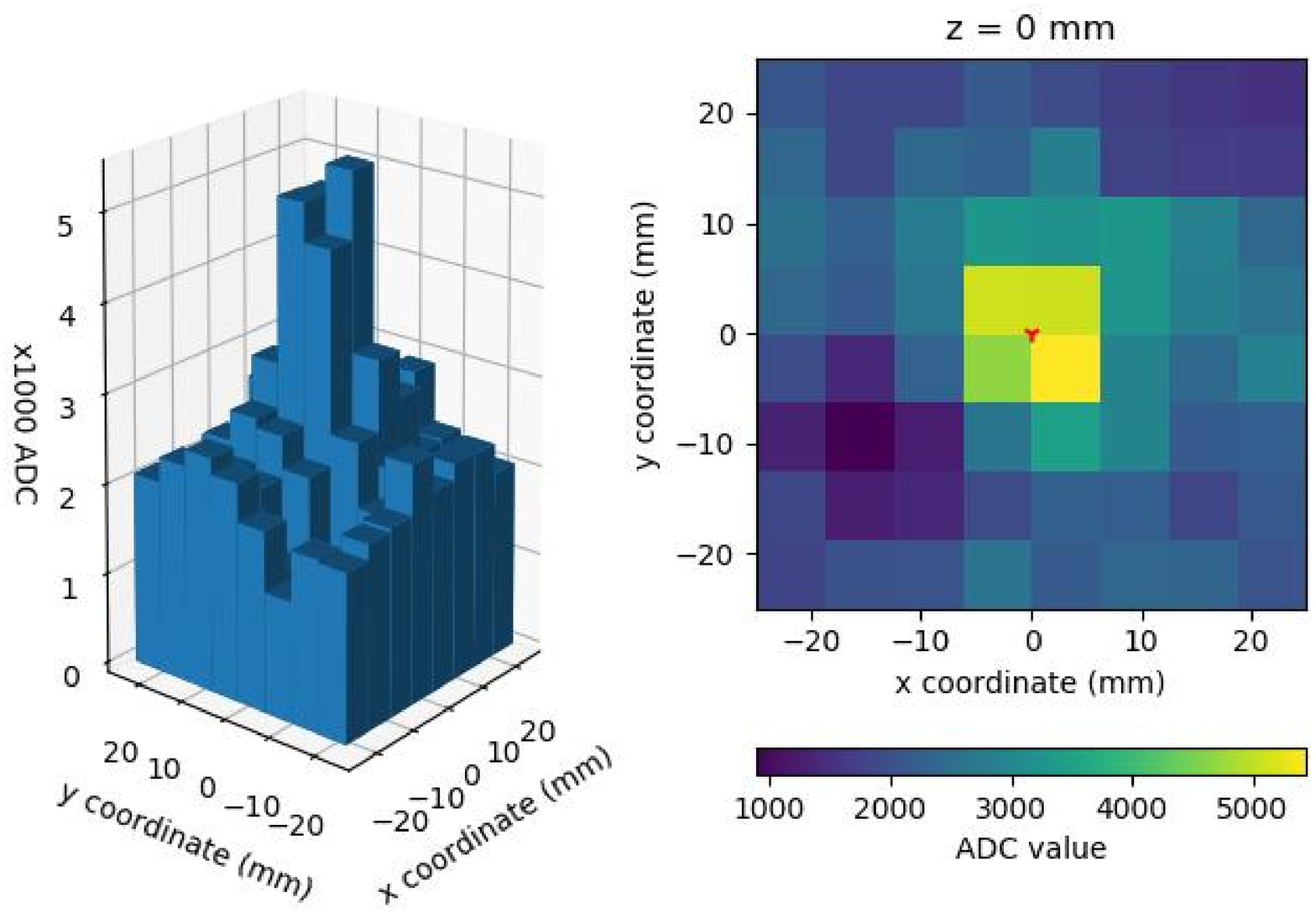}
\end{center}
\caption{Example of a light distribution of an event recorded by the module. Both parts of the figure show the same data, one in parallel perspective and the other in colorscale.}
\label{fig:good}
\end{figure}

\begin{figure}[htbp]
\begin{center}
\includegraphics[width=.9\linewidth,keepaspectratio]{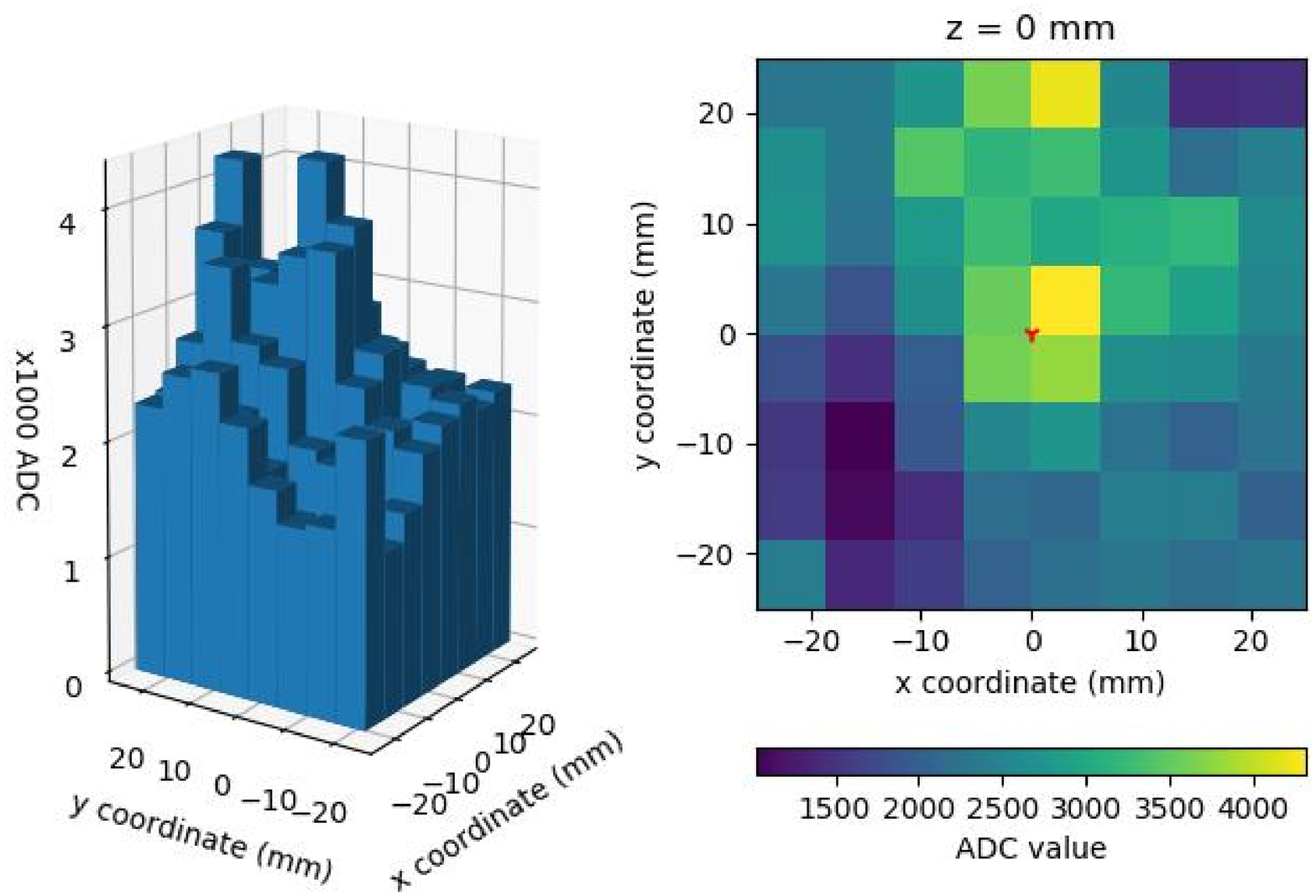}
\end{center}
\caption{Example of a light distribution composed of several energy deposits recorded by the module.}
\label{fig:compton}
\end{figure}

We define an event as one or several interactions of a single gamma ray that deposits enough energy to trigger the read--out electronics. Figure~\ref{fig:good} shows the scintillation light distribution of an event. Because of the dominant physics processes at these energies, gamma rays may interact several times {\vt by Compton scattering} in the scintillator. This leads sometimes to an event with several distinct energy deposits, as shown on Figure~\ref{fig:compton}.

These events can be identified automatically using an algorithm that checks whether the pixels that received the more scintillation light are packed together or separated. It relies on a 2D convolution of a shape representing the maximum authorized spread of the event with the scintillation light distribution. 

This analysis showed that about $15\%$ of events have a scintillation light distribution that is either spread or multiple. 

The reconstruction of the first interaction position is done by machine learning algorithms that analyze the shape of the scintillation light distribution. Events with a spread or multiple scintillation light distribution {\vt are less well reconstructed by the algorithms, and they may or may not} be used for position reconstruction depending on the intensity of the astrophysical source and the targeted sensitivity and angular resolution of the telescope. 

\subsection*{Artificial neural networks}

An artificial neural network (ANN) is a network of simple algorithms that aims at mimicking the operation of a biological neuron. 
Those algorithms execute a weighted sum of several inputs, and apply to this number an $\mathbb{R} \to \mathbb{R}$ function called activation function. We use ANNs to reconstruct the position of interaction of a gamma ray {\vt from the measured} scintillation light distribution. 

The machine learning used in this study is supervised learning. It means that ANNs are first trained to output the correct coordinates using data with known output, by minimizing an error function. Once the training is completed, reconstructing unknown events consists in executing the neural network program with the parameters calculated during the training process. The weights of the neurons are free parameters adjusted during the training of the ANN. 

ANNs used in this work are multi--layer perceptrons: neurons are organised into one or several layers, that are executed sequentially. The number of layers and their size (number of neurons) are meta--parameters of the ANN, chosen by the programmer. The activation function is also a meta--parameter, as well as the algorithm used for training. All this work has been done using the python Keras framework~\cite{keras} with Theano~\cite{theano} backend.

\subsection*{2D position reconstruction}

To check the effect of varying meta--parameters on the performance of neural networks, a systematic exploration of the meta--parameters space has been conducted. The analysis {\vt has been performed} for the following set of meta--parameters:
\begin{itemize}
    \item $0$ to $4$ hidden layers
    \item $6$ to $35$ neuron per hidden layers
    \item \verb|sigmoid|, {\vt \verb|hard sigmoid|,} \verb|tanh|, \verb|elu|, \verb|relu|, \verb|softplus|, \verb|softmax|, \verb|softsign|, and \verb|linear| activation functions~\cite{keras}
    \item \verb|adam|, \verb|nadam| and \verb|adamax| training algorithms 
\end{itemize}

\begin{figure}[tbp]
\begin{center}
\includegraphics[width=.99\linewidth,keepaspectratio]{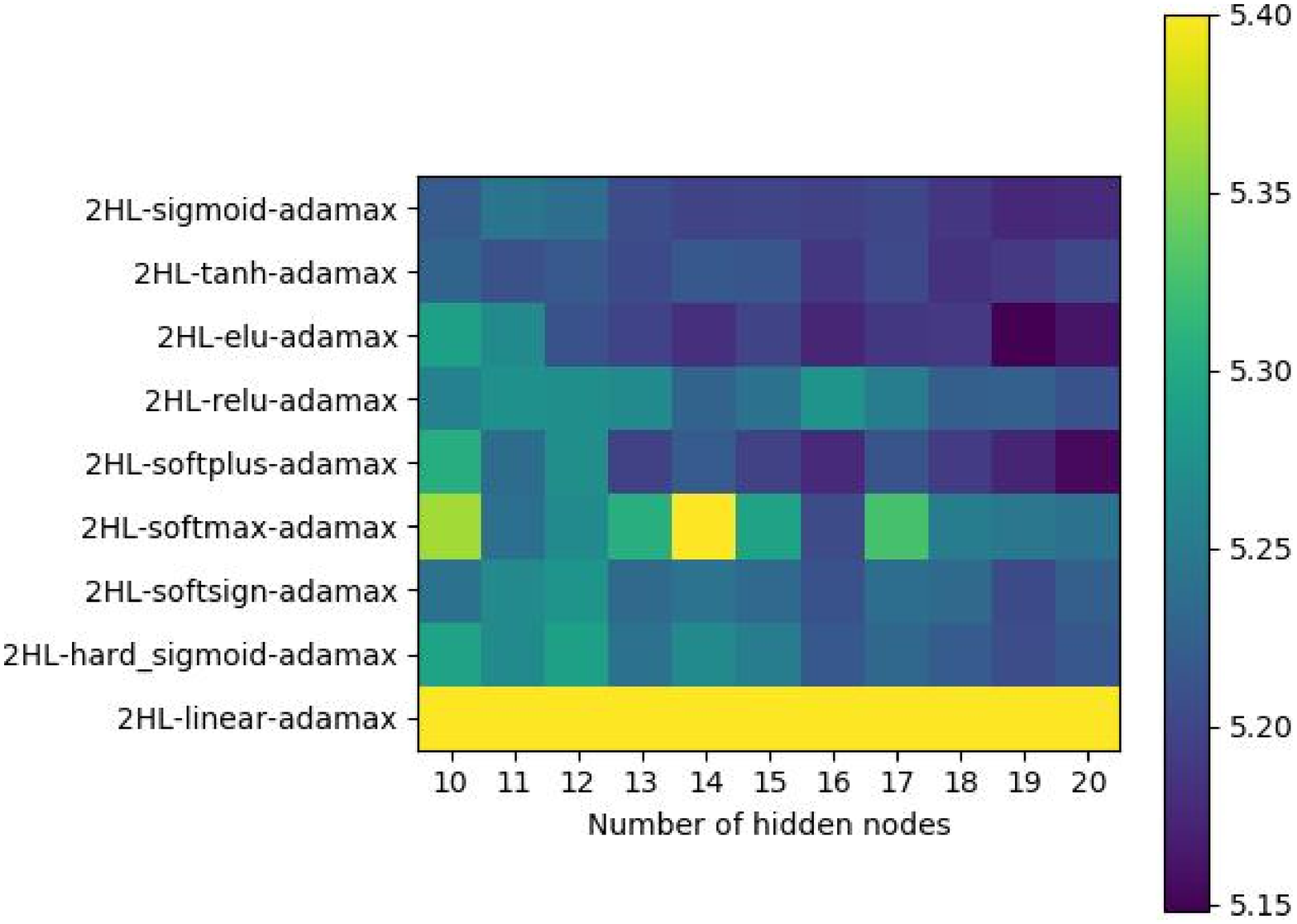}
\end{center}
\caption{Output example of a systematic meta--parameters space exploration. On the $x$ axis is represented the number of neurons per layer. On the $y$ axis are indicated {\vt the nine tested activation functions. Here the ANN uses two hidden layers (HL) and the adamax training algorithm}. Colorscale represents the performance of the ANN $\sigma_\mathrm{XY}$ as defined in the text (lower is better).}%\st{the number of layers (HL stands for hidden layer), the activation function and the training algorithm used} 
\label{fig:nn}
\end{figure}

Figure~\ref{fig:nn} shows an output example of such an exploration. {\vt %\st{The two axes represent a set of meta--parameters, and the} 
The colorscale represents} the performance of an ANN {\vt %\st{, $\sigma_\mathrm{XY}$. It is}, 
which is defined as} the square root of the mean squared distance from the reconstructed coordinates to the known ones,

\begin{equation}
    \sigma_{\mathrm{XY}} = \sqrt{ \frac{1}{N} \sum_{i=1}^N \left[ \left( x_i^\mathrm{known} - x_i^\mathrm{rec} \right)^2 + \left( y_i^\mathrm{known} - y_i^\mathrm{rec} \right)^2 \right]}
    \label{eqn:sigmaxy}~,
\end{equation}
where $ x_i^\mathrm{known} $ (respectively $ y_i^\mathrm{known} $) is the known $x$ (resp. $y$) coordinate of event $i$ and $ x_i^\mathrm{rec} $ (resp. $ y_i^\mathrm{rec} $) is the $x$ (resp. $y$) coordinate reconstructed using the ANN. {\vt Equation~\ref{eqn:sigmaxy}} is calculated on a set of $N=14450$ events evenly distributed on the $xy$ plane of the detector.

%vtTo check the effect of varying meta--parameters on the performance of neural networks, a systematic exploration of the meta--parameters space have been conducted. Figure \ref{fig:nn} shows an output example of such an exploration. The two axes represent a set of meta--parameters, and the colorscale the performance of an ANN, $\sigma_\mathrm{XY}$. It is the square root of the mean squared distance from the reconstructed coordinates to the known ones. It is calculated on a set of $N=14450$ events evenly distributed on the $xy$ plane of the detector, with the equation \ref{eqn:sigmaxy} where $ x_i^\mathrm{known} $ (respectively $ y_i^\mathrm{known} $) is the known $x$ (resp. $y$) coordinate of event $i$ and $ x_i^\mathrm{rec} $ (resp. $ y_i^\mathrm{rec} $) is the $x$ (resp. $y$) coordinate reconstructed using the ANN.

%vt\begin{equation}
%vt    \sigma_{\mathrm{XY}} = \sqrt{ \frac{1}{N} \sum_{i=1}^N \left[ \left( x_i^\mathrm{known} - x_i^\mathrm{rec} \right)^2 + \left( y_i^\mathrm{known} - y_i^\mathrm{rec} \right)^2 \right]}
%vt    \label{eqn:sigmaxy}
%vt\end{equation}

%V\begin{figure}[htbp]

It should be noted that this definition of the performance of an ANN includes in the calculation some events from the background radioactivity. Their position of interaction is not in general that of the gamma rays from the source, leading to an apparent worsening of the ANN performance from $\sigma_{\mathrm{XY}} \approx 5.2\,\mathrm{mm}$ to $\sigma_{\mathrm{XY}} \approx 2.8\,\mathrm{mm}$. {\vt In measurements with the $^{137}$Cs source, the background amounts to 2\% of the total number of counts in the selected energy window.}

%vtThis analysis have been conducted for the following sets of meta--parameters:
%vt\begin{itemize}
%vt    \item $0$ to $4$ hidden layers
%vt    \item $6$ to $35$ neuron per hidden layers
%vt    \item \verb|sigmoid|, \verb|tanh|, \verb|elu|, \verb|relu|, \verb|softplus|, \verb|softmax|, \verb|softsign|, and \verb|linear| activation functions
%vt    \item \verb|adam|, \verb|nadam| and \verb|adamax| training algorithms
%vt\end{itemize}

{\vt The analysis} has been conducted for various event selection procedures. All of them have in common a fine selection of the energy of the events around the full--energy peak of the source. 
%All ANN were trained using a validation dataset and Keras \verb|early_stopping| handle to prevent overlearning. For the most promising parts of the meta--parameter space, the ANN 2D localisation error $\sigma_\mathrm{XY}$ presented in equation \ref{eqn:sigmaxy} has been split in a size--of--spot error and an offset error.

\begin{figure}[htbp]
\begin{center}
\includegraphics[width=\linewidth,keepaspectratio]{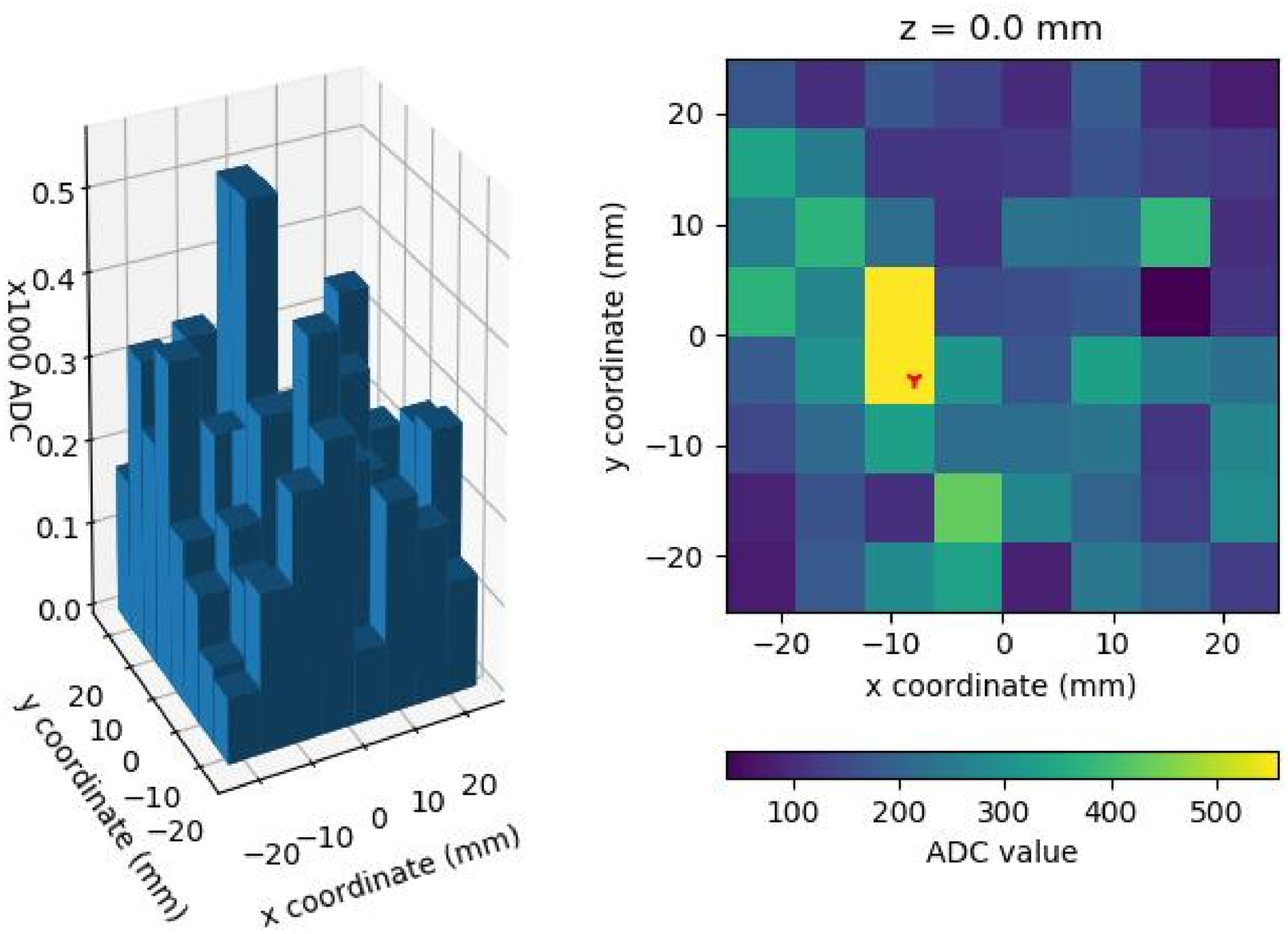}
\end{center}
\caption{Example of an event of energy $59.5\,\mathrm{keV}$ recorded by the module.}
\label{fig:241Am}
\end{figure}

Since the calorimeter must be able to detect and reconstruct the position of gamma rays of any energy, some datasets were acquired using the $59.5\,\mathrm{keV}$ line of a collimated ${}^{241}$Am radioactive source. An example of event at this energy can be seen on Figure \ref{fig:241Am}.

This study concluded that the most robust way of training an ANN {\vt for our application} is using \verb|adamax| and all event morphologies. %the least possible event selection procedures for the training dataset. Offset error is always smaller than size--of--spot error, showing that trainings have been efficient.

Since there is very little improvement {\vt with more than} $2$ hidden layers and $20$ neurons per hidden layer and since those numbers have an effect on the execution time of the reconstruction algorithm, these numbers have been found to be a good compromise. Considering the results shown on Figure~\ref{fig:nn} amongst others, we choose to use either the \verb|elu| or \verb|softplus| activation function. Finally, this study showed that using two neural networks, each trained on a single $59.5\,\mathrm{keV}$ or $662\,\mathrm{keV}$ energy window, does not perform significantly better than using a single neural network trained on both energy windows, with $\sigma_{\mathrm{XY}} \approx 2.8\,\mathrm{mm}$.

%Finally, this study showed that the performance of neural networks trained for specific energy windows is {\vt \st{identical to the performance} is not better than the one} of a neural network trained with both $59.5\,\mathrm{keV}$ and $662\,\mathrm{keV}$ energy windows.

\subsection*{Depth of interaction reconstruction}

The depth of interaction reconstruction algorithm has been designed using the knowledge acquired from the {\vt reconstruction of the $xy$ position}. The principle driving the position reconstruction along the $z$ axis is {\vt %\st{scintillation light diffusion in the CeBr$_3$ crystal: for a same energy deposit in the crystal, the closer it is to the photodetector, the more narrow it will be.} 
that the measured scintillation light distribution is expected to be narrower for gamma--ray interactions occurring closer to the photodetector.}
However, this effect is faint and therefore requires a deeper neural network. For that purpose we choose to focus the optimization on a $4$ hidden layers, $80$ neurons per layer ANN and optimize only the activation function and the training algorithm.

To train the neural networks, data was acquired by irradiating the $yz$ plane of the detector with a $^{137}$Cs radioactive source. These trainings rely only on the $z$ coordinate. To generate events of different energies, four energy ranges were defined, one in the full--energy peak at $662\,\mathrm{keV}$, the three others in the Compton front around $60\,\mathrm{keV}$, $130\,\mathrm{keV}$ and $300\,\mathrm{keV}$. These last three types of events approximate a localised full--energy event of a lower energy gamma--ray by the scintillation light of a Comptonized electron.

To assess the performance of these ANNs, we use {\vt the standard deviation} %\st{$\sigma_Z$ as defined in equation (2) using the same terminology as in equation (1)}}
\begin{equation}
    \sigma_{\mathrm{Z}} = \sqrt{ \frac{1}{N} \sum_{i=1}^N \left( z_i^\mathrm{known} - z_i^\mathrm{rec} \right)^2}~,
    \label{eqn:sigmaz}
\end{equation}
{\vt where the terminology is the same as in Equation~\ref{eqn:sigmaxy}.}

This part of the study concluded that ANNs are best trained for our purpose using \verb|adamax|. The most performing activation functions {\vt in this case} are \verb|elu| and \verb|relu|. The energies at which depth of interaction is best reconstructed are {\vt around} $662\,\mathrm{keV}$ and $300\,\mathrm{keV}$, with $\sigma_{\mathrm{Z}} \approx 2\,\mathrm{mm}$. In the $130\,\mathrm{keV}$ energy range, ANNs are significantly worse with $\sigma_{\mathrm{Z}} \approx 2.4\,\mathrm{mm}$, and in the $60\,\mathrm{keV}$ energy domain they perform even worse with $\sigma_{\mathrm{Z}} \approx 2.6\,\mathrm{mm}$. This could be explained by the shape of the scintillation light distribution at lower energies, as shown in Figure~\ref{fig:241Am}: The width of such a distribution is hard to measure and its peak is close to the noise level. For these lower energy gamma rays, the attenuation length in CeBr$_3$ is $< 2\,\mathrm{mm}$, therefore using a neural network of larger uncertainty is sub--optimal. Assigning to each of these events a depth of interaction equal to the mean depth of interaction of gamma rays with the same energy (calculated from NIST XCOM cross--section database~\cite{nistxcom}) gives $\sigma_{\mathrm{Z}} < 2\,\mathrm{mm}$.

Finally, as for the 2D position reconstruction, training ANNs with all events from the $662\,\mathrm{keV}$, $300\,\mathrm{keV}$ and $130\,\mathrm{keV}$ energy domains led to $\sigma_{\mathrm{Z}}$ results similar to {\vt those obtained with specific ANNs trained with events from each energy domain}. 
%{\vt Below $130\,\mathrm{keV}$, we assign a constant depth of interaction to the events between $0$ and $2\,\mathrm{mm}$ depending on the gamma-ray energy. \st{A well--chosen constant depth of interaction can be assigned to all events}}. 

\section*{Conclusion{\vt s}}

We developed a calorimeter module for next generation MeV--range gamma--ray observatories. {\vt We focused on reconstructing as best as possible the 3D position of the first impact of an incident gamma--ray within the detector, as this property is vital for the performance of a Compton telescope.} We chose for that purpose deep learning algorithms. 
2D position reconstruction algorithms were optimized to attain an uncertainty $\sigma_{\mathrm{XY}} \approx 2.8\,\mathrm{mm}$ using $2$ hidden layers, $20$ neurons per layer, \verb|elu| or \verb|softplus| activation function and \verb|adamax| training algorithm. 3D position is obtained by reconstructing the depth of interaction with a separate neural network using $4$ hidden layers, $80$ neurons per layer, \verb|elu| or \verb|relu| activation function and \verb|adamax| training algorithm. The uncertainty of this reconstruction is $\sigma_{\mathrm{Z}} \approx 2\,\mathrm{mm}$. Those results apply to any gamma--ray energy from $60\,\mathrm{keV}$ to $662\,\mathrm{keV}$.

{\vt Finally, using an energy correction dependant on the position of interaction \cite{2016NIMPA.832...24G}, the spectral resolution of the module is found to be $5.1\%$ at $662\,\mathrm{keV}$.}